\documentclass{aa}

\usepackage{txfonts}
\usepackage{graphicx}
\usepackage[mathcal]{eucal}
\usepackage{amssymb}
\usepackage{natbib}

\begin{document}

\title{A quantitative analysis of stellar activity \\ based on CoRoT\thanks{The CoRoT space mission, launched on 2006 December 27, was developed and is operated by the CNES, with participation of the Science Programs of ESA, ESA's RSSD, Austria, Belgium, Brazil, Germany and Spain.} photometric data}

\author{J.C. Hulot\inst{1}\and F. Baudin\inst{1}\and R. Samadi\inst{2}\and M.J. Goupil\inst{2}}

\institute{	Institut d'Astrophysique Spatiale, CNRS/Universit\'e Paris XI
		UMR 8617, F-91405 Orsay, France
\and 		LESIA, UMR8109, Universit\'e Pierre et Marie Curie, Universit\'e
		Denis Diderot, Observatoire de Paris, 92195 Meudon, France}

\date{Received ; Accepted }

\offprints{J.C. Hulot}

\abstract
{The CoRoT satellite has made available high precision photometric observations \textbf{of} a large number of stars \textbf{of} different spectral types. Continuous photometric time series allow the characterization of stellar microvariability in a systematic way.} 
{\textbf{We} determine an index indicating the level of activity, derived from photometric data, for a large sample of stars with different color temperatures. \textbf{We also} assess to \textbf{what} extent this index can be related to an estimated Rossby number \textbf{for stars whose rotation period can be estimated from the light curve}. We also estimate a characteristic \textbf{life}time of \textbf{the} surface heterogeneities causing the variability of selected light curves.}
{\textbf{Most work on the photometric impact of stellar activity has been based either on measuring the variance of stellar microvariability, or on fitting the light curve. Following similar research for the Sun, our work is based on the Fourier analysis of stellar light curves.} We have analyzed the Fourier power spectra of 430 selected light curves obtained by CoRoT during three observation runs. The low-frequency contribution of the stellar variability is modelled by a ``generalized semi-lorentzian'' profile. An activity index is derived from the fitted amplitude and width of the semi-lorentzian model. Some of the Fourier spectra \textbf{exhibit} a rotational modulation \textbf{which enables} the determination of the rotation period. In addition, a convective turnover time is derived from a grid of stellar models, so that a Rossby number can be estimated for this subsample of stars. A characteristic \textbf{life}time of the phenomena causing the observed power at low frequency is \textbf{assessed} from the fitted model of the power spectrum and is compared to the rotation period.}
{Higher values of the microvariability index are observed among the coolest stars from our sample. 28 light curves show a clear rotational modulation. The rotation periods derived from the observed low-frequency peaks in the Fourier spectrum decrease with the color temperature, in accordance with previous observations. The estimated Rossby number of most of the observed stars with a rotational modulation is \textbf{less} than 1, \textbf{generally accepted} as a critical value under which stars are expected to be active. \textbf{The activity index, computed from the Fourier spectrum,} decreases with increasing Rossby number.
%Original results are obtained in relation with the characteristic evolution time of the phenomena causing the observed rotational modulation.
The quality of the CoRoT data enables the determination of the characteristic \textbf{life}time of active structures. \textbf{It is} shown to increase with the rotation period, \textbf{but some scatter could arise from different surface heterogeneities (spot and faculae for example)}.% and activity mechanisms.
}
{}

\keywords{stellar microvariability, stellar activity}

%\authorrunning{Hulot et al.}
\titlerunning{Stellar activity from CoRoT data}

\maketitle

\section{Introduction}
\label{sec:intro}

Since \citet{Kron1947,Kron1950} suggested that spots on the surface of \textbf{stars other than the Sun} could explain their photometric variability, much evidence for stellar activity has been collected. Activity refers here to a complex set of phenomena, including spots, faculae, eruptions, which have been observed on the Sun and are now suspected to exist on other stars. Activity is generally considered to have a magnetic origin. However, the physical mechanisms that explain the observed active structures and their {\bf temporal} evolution are still poorly understood. Observations of the Ca emission lines allowed the identification of \textbf{stars with chromospheric activity, some of them having an activity cycle} \citep{Wilson1978}. Some time series of Ca lines fluxes show a rotational modulation, which is useful for determining the rotation period \citep{Vaughan1981,Baliunas1983}, \textbf{as for the Sun}. Co-rotating \textbf{photospheric} heterogeneities, such as stellar spots, can \textbf{also} result in broad-band photometric microvariability. \textbf{Large enough spots} may be detected through the rotational modulation they induce in the star light curve. Due to their intrinsic temporal evolution, the surface heterogeneities are also expected to produce a power excess at low frequency in a Fourier spectrum of a \textbf{velocity} or photometric time series \citep{Harvey1985}. These heterogeneities can be considered as a random process with some memory. \textbf{They give rise to} an exponentially decaying autocorrelation in time, yielding a Lorentzian shape in the power spectrum. In solar observations, at low frequency, several processes contribute to the power spectrum: supergranulation \citep[in velocity series,][]{Jimenez1988}, bright points \citep[in photometry,][]{Harvey1993} or granulation (in both). As it is the process with the longest time scales, magnetic activity seen through the signature of transiting spots on the solar surface in photometric series is detected at the lowest frequencies in the power spectrum. \textbf{In solar-type stars, we expect the active structures to be the phenomena with the longest time scales. \citet{Mosser2009} have shown that, in the case of several solar-type stars, the microvariability at the lowest frequencies is well modelled by the transit of stellar spots. In addition, simulation of the granulation and comparison with observations in the case of one of these solar-type stars (HD49933) shows that granulation has time scales shorter than that of spots \citep{Ludwig2009}, as for the Sun. Modeling of the spectrum with Harvey profiles is commonly used in the seismic analysis and shows a similar shape to that of the Sun \citep{Michel2009,Benomar2009}. It is not excluded that other physical processes, such as large convective cells for instance, which are suspected to exist in giant stars, could have the longest times. However, we selected our targets among stars that are identified as main-sequence}.

\textbf{Stellar activity} has been \textbf{raised} as an obstacle to the detection of exo-planets around cool stars (e.g. \citet{Aigrain2004}, \citet{Meunier2010a}, \citet{Meunier2010b}). \textbf{The impact of stellar activity on exoplanet detectability has been estimated based on the solar case and on scaling laws \citep{Aigrain2004}. Simulations have given estimates of the ``stellar activity noise'' that strongly depend on spectral types. Actual light curves collected by the CoRoT (Convection, Rotation and planetary Transits) now provide us new insights on the microvariability that can be attributed to stellar activity.}   

As stressed by \citet{Ribarik2003}, CoRoT gives the opportunity to analyze a large amount of photometric data related to spotted stars. While ground-based observations have a limited precision and a strongly constrained observational window, the space-based CoRoT photometric instrument has collected numerous high-precision and continuous light curves. It has been designed to study several bright seismic targets and to search for exoplanets around a large amount of fainter stars. Seismic and exoplanet targets belong to different spectral types and luminosity classes. Exoplanet targets have however been mainly selected \textbf{from} F to M spectral types. \textbf{The observed noise from an exoplanet search point of view has been measured and has been shown to be from two to three times the expected photon noise \citep{Aigrain2009}. We have performed a more detailed analysis of the stellar microvariability based on the Fourier power spectrum, which allows us to discriminate several phenomena with different time scales.} The purpose of this work is to show, based on a still limited but homogenous sample, that the available light curves {\bf can} deepen our knowledge of stellar activity as the time scales of magnetic structures provide hints to understand the underlying mechanisms. 

There are at least two approaches to process the thousands of available light curves. The first is based on directly modeling a time series, as suggested by \citet{Ribarik2003} and applied by \citet{Lanza2009},  \citet{Mosser2009} and \textbf{\citet{Fröhlich2009}}. Models of light curves are based on various assumptions regarding the number of spots, their physical properties, differential rotation and limb darkening. There is no unique solution, but some ``reasonable'' hypotheses can give interesting results for a case by case analysis. A second approach is based on the Fourier power spectrum of a time series \citep{Harvey1985}. It can be more reliable in identifying different features \textbf{in a given signal as it can discriminate different  time scales and quantify the lowest frequency contributions to the stellar microvariability. In addition, it allows the identification, in some cases, of a rotational modulation}. This second approach is adopted in the present work. Our work is based on a 430-star sample, which is large enough to obtain several interesting results regarding stellar activity. Here we develop and test an analysis method on the present sample in order to determine physical constraints on stellar activity.

In Section~\ref{sec:sample}, we present our initial sample of 430 stars and describe our analysis method of the Fourier power spectra of the selected light curves. In Section~\ref{sec:microvar}, we first {\bf summarize} some useful results related to the solar microvariability.
%\textbf{Different time scales have been already attributed respectively to activity, supergranulation, granulation and noise. Similar contributions are expected to be observed in solar-like stars. The solar power spectrum has been already modeled as the sum of several generalized lorentzian contributions. The same kind of power spectra has been simulated for estimating the ``stellar activity noise'' onto the exoplanet detectability. A generalized semi-lorentzian profile has been already used in several works on stellar oscillations modes \citep{Benomar2009} and stellar granulation \citep{Ludwig2009}}.
We then define \textbf{our stellar activity index computed from the Fourier spectrum.} The distribution of \textbf{ our activity index} against the color temperature is analyzed. In Section~\ref{sec:rotmod}, we \textbf{present our findings related to} 28 light curves\textbf{, out of the 430-star sample,} showing a rotational modulation. In Section~\ref{sec:Rossby}, the Rossby number for stars showing a rotational modulation {\bf is} derived. The correlation between the observed \textbf{ activity index} and the estimated Rossby number is investigated. In Section~\ref{sec:evtime}, we analyze our results regarding the characteristic evolution times of microvariability. In Section~\ref{sec:conclu}, we conclude and discuss the opportunity to go further into a wider analysis of the CoRoT data.

\section{The sample of CoRoT light curves}
\label{sec:sample}

\subsection{Selection criteria} 
\label{sec:selec crit}

Our sample\footnote{All data are available from the public CoRot archive at: {\tt idoc-corot.ias.u-psud.fr}} was initially made of the first 30 seismic targets observed by CoRoT during three observation ``runs'' (IRa01 --January/April 2007-- SRa01 --April/May 2007-- and LRc01 --May/October 2007) and of the 400 brightest exoplanet targets observed during the initial run (IRa01). The apparent V magnitudes of the seismic targets range from 5.45 to 9.48, \textbf{and} between 12 and 13 for the selected exoplanet targets. Our sample includes stars with spectral types from B to M, but most of the exoplanet targets are redder than A. Most of the selected light curves are 60-days long. However, some of the seismic targets light curves are limited to 30 days and a few are 150 days long. Seismic targets are observed in a broad-band white-light flux. For exoplanet targets, the flux is the sum of three broad-band chromatic fluxes, equivalent to a white-light flux. There is no bolometric correction. Broadband photometric data are collected with two sampling times (32 or 512 s) depending on the kind of target. The available photometric time series {\bf have} a very high precision that allows, after correcting for an instrumental linear trend with time, a systematic study of stellar microvariability. More details on the CoRoT time series \textbf{are given by} \citet{Auvergne2009}.

\subsection{Available stellar information} 
\label{sec:stell info}

Apparent B, V, R and I magnitudes were determined during ground-based photometric campaigns in preparation {\bf for} the CoRoT mission \citep{Deleuil2009}. A color temperature was derived from those magnitudes, which were not corrected for interstellar extinction \citep{Deleuil2009}. We checked that there is a monotonic relationship between this color temperature and the observed B-V index of the selected stars. In the next sections, we use this temperature as a proxy {\bf for} the stellar spectral type.

\subsection{Analysis procedure}
\label{sec:method}

\begin{figure}
\centering
\includegraphics[height=6cm,angle=90]{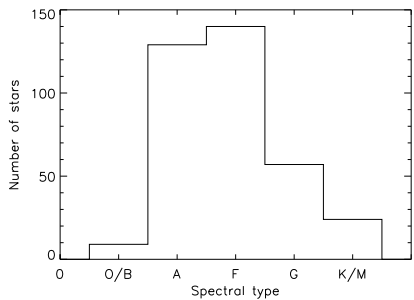}
\caption{The distribution of the selected stars by spectral type.}
\label{histotype}
\end{figure}

Each light curve has been analyzed through a 4-step process. The first step starts with a visual inspection of the light curve. It is performed in order to identify any significant discontinuit{\bf ies} of the recorded flux which could be caused by a high-energy particle on the photometric detector. \textbf{As the resulting discontinuities do not show any systematic time scale, no automatic procedure worked well enough to remove their impact in the time series or in the Fourier power spectrum. Consequently, a visual inspection has been performed \citep[for more details on these discontinuities, see][]{Auvergne2009}}. The light curves showing such discontinuities followed by a slow {\bf recovery} are excluded from the next steps of our analysis. Exceptions are accepted if the discontinuity occurs close to the extremities of the time-series, so that the truncated time-series, excluding the discontinuity, is long enough to be included in the further analysis. All {\bf of} the seismic targets went through the first step, but 67 exoplanet targets had to be excluded. \textbf{A linear variation was removed} with an ordinary least square linear fit. The observed temporal drift is mainly attributed to instrumental aging. We checked that a model with a second or third order polynomial drift does not significantly change the Fourier power spectrum. {\bf Missing} observations are replaced by linearly interpolated values. {\bf They account for} close to 10 \% {\bf and} mainly result from the crossing of the South-Atlantic Anomaly by the CoRoT spacecraft. \textbf{Their signature appears in the Fourier spectrum at the orbital frequency and harmonics.} The seismic targets include 4 stars that were found out to be binaries. They were excluded from our analysis. At the end of this first step, our sample {\bf includes} 359 stars (Table~\ref{tab:steps}) which are mainly distributed within the A to G spectral types (Figure~\ref{histotype}).

%\begin{table}
%\centerline{\begin{tabular}{|c|c|}
%\hline spectral type & number\\
%\hline O/B  & 9 \\
%\hline  A & 129 \\
%\hline F & 140 \\
%\hline G & 57 \\
%\hline K/M & 24 \\
%\hline Total & 359 \\
%\hline
%\end{tabular}}
%\caption{Number of selected light curves by spectral types at the end of the first step of our analysis method}
%\label{tab:nobs}
%\end{table}

In a second step, we derive the Fourier power spectrum from a uniformly sampled time series. The third step consists in computing several indexes to characterize the low-frequency power spectrum. Narrow spectral peaks are searched for and their frequencies are compared in order to identify the rotation period (see Section \ref{sec:rotmod}). The fourth step is designed to fit a generalized semi-lorentzian model to the low frequency spectrum. The power spectra of 58 stars, among the hottest ones from our sample, were however too flat to be fitted to our model. The corresponding stars were probably not active or had too low activity to be detected. The number of stars {\bf surviving} the different steps of our analysis is shown in Table~\ref{tab:steps}.

\begin{table*}
\centerline{\begin{tabular}{|c|c|c|c|}
\hline step & number of   & number of  & total number  \\
 & seismic targets  & exoplanet targets & of stars \\
\hline initial selection  & 30 & 400 & 430 \\
\hline  after excluding 'hot pixels' and binaries & 26  & 333  & 359 \\
\hline fitted Fourier power spectrum & 25  & 276  &  301 \\
\hline with rotational modulation & 6  & 22  &  28 \\
\hline with estimated Ro  &  6   &  21  &  27 \\
\hline
\end{tabular}}
\caption{Number of selected light curves through the different steps}
\label{tab:steps}
\end{table*}

\section{Low-frequency microvariability}
\label{sec:microvar}

\subsection{The solar case}
\label{sec_solcase}

The solar microvariability has been extensively analyzed based on the SOHO/VIRGO  photometric data \textbf{\citep{Aigrain2004,Sel2003} before being extrapolated to other stars (using scaling laws to describe the rotation, the amplitude and characteristic evolution time of activity, in order to simulate an ``activity noise'' impact on exo-planet detectability} by \citet{Aigrain2004}. VIRGO data can be compared to CoRoT data, as they are also intensity measurements. However, the effects of the instruments (bandwith) are different and should be taken into account. VIRGO data are based on 3 narrow bandwidths, whereas CoRoT relies on one wide band. \textbf{As shown by} \citet{Aigrain2004}, VIRGO power spectra show several contributions which are attributed to solar activity, super-granulation and granulation. The first contribution is observed at the lowest frequencies and shows a clear variability {\bf through} the solar cycle. A lorentzian profile was first suggested by \citet{Harvey1985} to model the low-frequency background in such a power spectrum. Each contribution to the power spectrum is actually better described by a 'generalized' lorentzian profile, the total profile being the sum of each contribution:

 \begin{equation}
 P(f) =  \displaystyle{\sum_{i=1}^{N}}  \frac{A_i}{1+\left(\frac{f}{\Delta f_i}\right)^{\alpha_i}} + B
 \label{eqn:power_density}
 \end{equation}

$P(f)$ is the power density in ppm$^2 \mu$Hz$^{-1}$, {\bf with ppm for part-per-million.} $\alpha_i$  equals {\ 2} when the phenomenon causing the flux variation exponentially decays, but different values can result from more complex temporal evolution. $A_i$ is the amplitude of the i-th contribution and $1/\Delta f_i$ is its characteristic time. $B$ corresponds to an additional white noise. It is important to stress that, in the solar case, the low frequency contribution has a 10-day characteristic time \textbf{\citep{Aigrain2004}}. It is then significantly shorter than the evolution time of a solar active region, generally estimated around 2 months, and should be interpreted as a characteristic evolution time of individual spots \textbf{and faculae} forming an active region. 
%When fitting the VIRGO data to the above model, we have found that the amplitude $A_1$ of the solar activity contribution varies by two magnitudes from activity minimum to maximum. \textbf{This result confirms those previously obtained by \citet{Aigrain2004} on a shorter time series.} 

\subsection{A simplified model of the low-frequency stellar microvariability}

A variance derived directly from a light curve could capture different superposed phenomena. On the contrary, a Fourier analysis allows the identification and quantification of the low-frequency background contribution. What is observed in the solar case is expected to be seen in \textbf{some} stellar light curves if the photometric precision is high enough. The low-frequency part (below 100\,$\mu$Hz) of the power spectrum may result from several contributions that must be disentangled. A first contribution to the power excess at low frequency \textbf{results from} instrumental effect\textbf{s}. All our light curves are corrected for a linear trend and the CoRoT stability is confirmed by a significant number of flat power spectra, mainly observed for A stars. A second contribution is {\bf from} the so-called ``hot pixels'' that result from incident high-energy particles on the photometric detector. We recall that the light curves concerned are excluded from our analysis through a visual inspection, so that this contribution should not be observed in our sample. A third contribution to the power spectrum relates to intense and narrow peaks that are due either to long-period oscillations or to a rotational modulation. %Peaks of pulsating stars have quite large amplitudes and their frequencies generally range from 30 to 60\,$\mu$Hz. The peaks due to a rotational modulation are quite weak, and their fundamental frequency is below 10\,$\mu$Hz.
The fourth contribution is the one \textbf{that} we try to identify and to model. It corresponds to a low-frequency background and is well described  by a continuous profile centered on zero frequency (see Eq.~\ref{eqn:power_density}). It varies significantly from {\bf one} light curve to another. Our sample shows that an A-star spectrum is more often flat, while a G-star spectrum shows a higher power at low frequencies. \textbf{We suppose that 
late-type main sequence stars show surface heterogeneities with time scales that follow the same hierarchy as in the solar case, that is activity has a longer time scale compared to supergranulation and granulation. This relies on previous observations: first, as already mentioned, the results of \citet{Mosser2009} about spot modeling; second, in the case of HD49933, the power spectrum expected from a granulation model has been compared to the actual power spectrum observed with CoRoT and show the expected hierarchy of time scales \citep{Ludwig2009}. 
%Even if some differences still remain to be explained, the same kind of power spectrum and comparable time scales are found both with the theoretical model and in the observed light curve.
However, we cannot exclude that a giant star is included in our sample and that its low-frequency microvariability comes from another phenomenon than magnetic activity.
Such a case is likely to happen within some red giants whose surface could show large convective cells with long evolution time scales. Our sample of stars has been selected from main-sequence stars from the CoRoT database but misclassifications cannot be excluded.}

Considering the power spectrum at frequencies lower than 100\,$\mu$Hz, we exclude the peaks that may result from a rotational modulation or from long-period oscillations. This 'cleaned' power spectrum is then fitted to a \textbf{single} generalized semi-lorentzian distribution to which we add a constant free parameter:

 \begin{equation}
 P(f) =   \frac{A}{1+\left(\frac{f}{\Delta f}\right)^{\alpha}}+ \tilde{B} 
 \label{eqn:power_density2}
 \end{equation}

$A$ is the amplitude of the lower-frequency contribution to the \textbf{spectrum and $\Delta f$ is related to its characteristic time. It captures as shown in the solar case} \citep{Harvey1985}, the stellar activity. $\tilde{B}$ captures all the other contributions, including granulation, and not only the additional white noise as in Eq.~\ref{eqn:power_density}.
%In the case of a low-activity or even non-active star, it is not excluded that our generalized semi-lorentzian model captures granulation or supergranulation contributions. It should however be identified through the value of $1/\Delta f$. We come back to a more detailed interpretation of $1/\Delta f$ in Section~\ref{sec:evtime}.

\subsection{The low-frequency background index}

We derive a microvariability index from the generalized semi-lorentzian model (Eq.~\ref{eqn:power_density2}). We call it a ``low-frequency background index'' (LFBI). It is defined as  $I_{LF} = A \Delta f$ and it is expressed in ppm$^2$. It represents the integral of the power density of the model which describes the microvariability disentangled from other contributions. \textbf{A possible bias in the computation of the index corresponds to the case of a slowly rotating star whose light curve has an unresolved rotational modulation. In such a case, the rotational modulation peak cannot be removed from the power spectrum: it affects the very first bins of the spectrum and may lead to an over-estimated LFBI.}
In order to use the solar case as a reference when considering CoRoT light curves, \textbf{we have analyzed some} VIRGO data. \textbf{These data} need to be corrected for the instrument bandwith (see Section\,\ref{sec_solcase}), following \citet{Michel2009}.The correction factor \textbf{to VIRGO data to be compared with CoRoT data} is estimated to be 1.4. The results obtained from the VIRGO solar data at different times of the solar cycle indicates that  $A \Delta f$  is a representative index for activity.
%at least when stellar activity includes photospheric heterogeneities.
In the solar case, while $I_{LF}$ is close to 1.4\,10$^5$\,ppm$^2$ at {\bf the} maximum of activity, it is nearly 100 times lower at its minimum. \textbf{Correspondingly, $A$ varies from 1.5\,10$^5$\,ppm$^2 \mu$Hz$^{-1}$ at maximum to nearly 100 times lower at minimum, while $\Delta f$ remains close to 1, which corresponds to nearly 10 days.}

%We remind that the power excess at low frequency may however capture other contributions. The solar power spectrum is, for instance, better described by at least three contributions, which are related respectively to the activity, the mesogranulation and the granulation. The first one is however the largest one at low frequency, i.e. below 10 $\mu$Hz.

In order to check the robustness of our index $I_{LF}$, two additional indices were computed, {\bf consisting} of the direct integration of the Fourier power spectrum, respectively from 0 {\bf up} to 10\,$\mu$Hz and from 0 {\bf up} to 100\,$\mu$Hz. They were found to be highly correlated with $I_{LF}$, but we recall that these latter indices take into account all {\bf of} the power at low frequency, whereas $I_{LF}$ does not take into account the power due to pulsation peaks, peaks due to rotational modulation or the constant term capturing other phenomena and noise.

\subsection{A study case: HD\,49933}

\begin{figure}
\centering
\includegraphics[height=6cm]{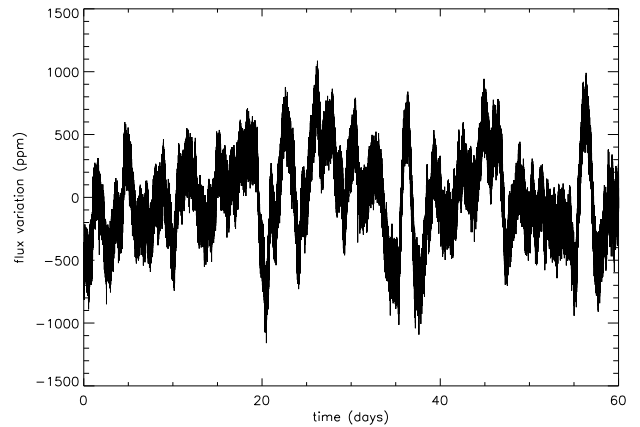}
\caption{HD\,49933 light curve. The observed signal has been corrected for a linear trend and converted into a relative flux variation and presented in ppm.}
\label{HDlight}
\end{figure}

\begin{figure}
\centering
\includegraphics[height=6cm]{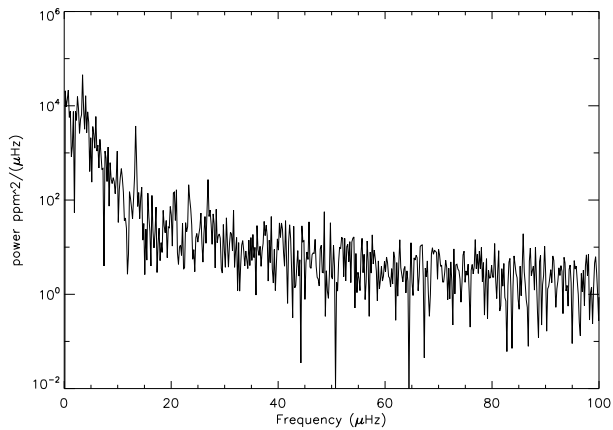}
\caption{Power spectrum of HD\,49933 computed from a 60-day time series after it was corrected for a linear trend which captures some instrumental effects. The power is presented here on a logarithmic scale \textbf{that allows us to see the rotational modulation peak at 3.5 $\pm$ 0.2 $\mu$Hz and its third harmonics.} .}
\label{HDspectrum}
\end{figure}

HD\,49933 is a bright solar-like star for which several parameters are now well known from ground-based observations. It is an F5 star with a visual magnitude 5.77. Its effective temperature is between 6500\,K and 6780\,K \citep{Ryabchi09,Bruntt2008,Bruntt2009}.
HD\,49933 was observed by CoRoT during its 60-day initial observation run with a 32s sampling time. The detrended stellar intensity shows some significant variations with a $\sim$1000 ppm amplitude (see Fig. \ref{HDlight}). We note that the VIRGO photometric data show a peak-to-peak variation which is also close to 1000 ppm at the maximum of solar activity. The power spectrum of HD\,49933 light curve confirms a power excess at low frequency (see Fig. \ref{HDspectrum}).

\begin{figure}
\centering
\includegraphics[height=6cm]{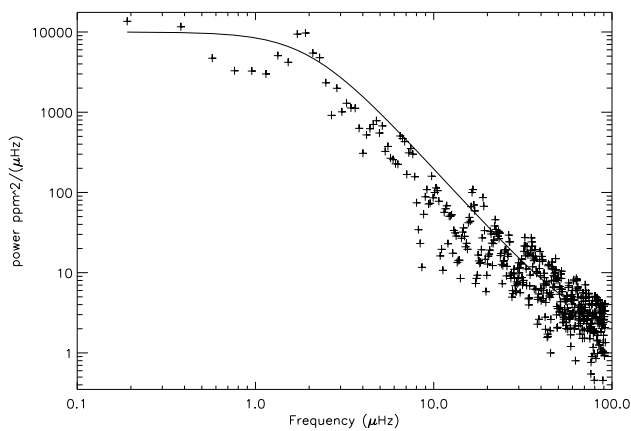}
\caption{A 'generalized' lorentzian model (with a uniform background) is fitted (solid line) to the power spectrum of HD\,49933 (crosses), from which the rotational modulation peaks have been removed.}
\label{HDfit}
\end{figure}

Following the method presented above, we obtained the following fitted parameters (see Fig.~\ref{HDfit}), \textbf{using a MLE fitting algorithm described by \citet{Appourchaux2008}}:  $A$\,=\,1.0\,10$^4$\,ppm$^2 \mu$Hz$^{-1}$, $B$\,=\,1.7\,ppm$^2 \mu$Hz$^{-1}$, $\Delta f$\,=\,2$\mu$Hz and $\alpha$\,=\,2.4. $A$ is about 15 times lower than the observed value for the Sun at its activity maximum, but less than 10 times larger than the same parameter at the Sun's minimum of activity. \textbf{The HD 49933 activity index is 2.\,10$^4$\,ppm$^2$.} Observations on a long time scale would be necessary to get a better knowledge of its activity cycle \citep{Garcia2010,Wilson1978}.

\subsection{Color temperature and microvariability}

%When plotted against the color temperature, the activity index shows a large scatter, particularly in the 5,000-7,000 K range (See Fig. \ref{index}). The same result is obtained with two other activity indexes defined in section 3.3.

%\begin{figure}
%\centering
%\includegraphics[height=6cm]{prodAf.ps}
%\caption{The low-frequency background power index, or microvariability index, in $ppm^2$, is plotted against the color temperature which is, within our sample, a monotonic function of the observed B-V index. There is a large scatter of the observed values, but higher values are obtained for the coolest stars.}
%\label{index}
%\end{figure}

\begin{figure}
\centering
\includegraphics[height=6cm]{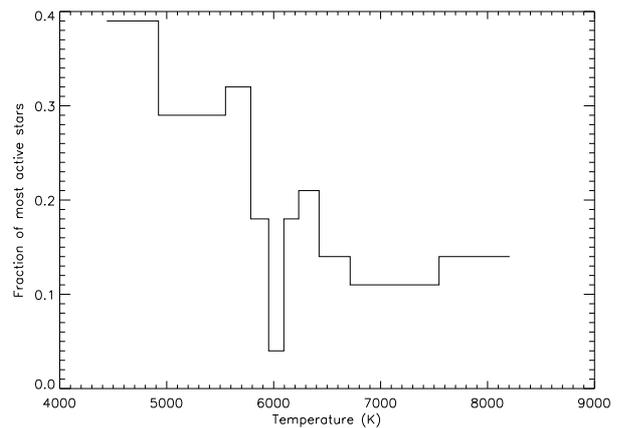}
\caption{The fraction of \textbf{most active} stars, \textbf{defined as stars} with a LFBI higher than a given threshold (see text). \textbf{Here, the threshold was defined so that the most active stars account for 20 \% of the total sample.} The \textbf{fraction} of stars with an index higher than the threshold is larger for the coolest stars. This \textbf{general trend} is not sensitive to the threshold value.}
\label{active}
\end{figure}

The LFBI was computed for the 301 stars of the sample for which a fit could be performed. \textbf{As was noted in section 2.3, 58 power spectra could not be correctly fit. All of them were ``flat'' spectra with a low power density at low frequency. Such spectra confirm the instrument stability. The corresponding stars are among the hottest and do not show a photometric signature of surface heterogeneities.} The activity indexes do not show a clear dependence on color temperature.
However, our sample is not uniformly distributed in terms of temperature: F stars are over-represented. In order to avoid this bias, we sort our sample by increasing temperatures, and divide it into nine sub-samples, each of them including nearly the same number of stars. We determine a threshold value of the LFBI defined {\bf so that} 20\% of the {\em full sample} has an index higher than the threshold. We then determine, within {\em each temperature sub-sample}, the fraction of stars with a LFBI higher than the threshold \textbf{and labelled by the fraction of most active stars.} The fraction of \textbf{most} active stars appears to be clearly related to the temperature, being higher in the low temperature sub-samples (see Fig. \ref{active}). As the 20\% threshold is arbitrary, we checked that the same result is observed when the threshold is determined so that we select respectively the 10\%, 30\%, 40\% or 50\% most active stars from the whole sample. The higher fraction of active stars among the \textbf{G} to K spectral types is consistent with previous observations \citep{Berdyugina2005}.
\textbf{According to our results, cool stars are} more likely to show an enhanced microvariability.

\section{Rotational signature}
\label{sec:rotmod}

\subsection{Rotational narrow peaks}

Some of the analyzed power spectra show super-imposed narrow peaks in addition to the low frequency background.
%More precisely, some high-power peaks at low frequency correspond to a fundamental frequency and to its harmonics.
Such peaks are attributed to the transit of surface heterogeneities which are large and contrasted enough to cause a periodic intensity modulation \textbf{over a significant fraction of the observation} \citep[see][]{Mosser2009}. A Fourier analysis shows that a single co-rotating permanent spot results in a strong peak and weaker even harmonics in the Fourier power spectrum of its photometric flux \citep{Clarke2003}. Indeed, we often observed harmonics above the peak corresponding to the rotation period. We checked with simulated light curves that a more complicated distribution of spots or, more generally of active areas, gives the expected peaks, with even and odd harmonics. These peaks are thus considered as the signature of activity {\em and} rotation. \textbf{Because the observation time is limited, the frequency resolution might be too low to detect such peaks in slowly-rotating stars. As previously noted, the LFBI of such stars could be over-estimated.}

\subsection{An example of rotational modulation : HD\,49933}

In order to identify a rotational modulation in the CoRoT light curves, we first focused on the brightest stars which were selected as seismological targets, and chose HD\,49933 among them. While p-modes were detected in this star, a rotational modulation of the observed spectral lines was attributed to one or more spots \citep{Mosser2005}. In order to illustrate how we carried out our analysis on the full sample of stars, we present in more detail our results for this star. 

The power spectrum in the range [0,20]\,$\mu$Hz shows a strong peak and weaker ones (see Fig.~\ref{HDspectrum}). The main peak is centered at 3.5 $\pm$ 0.2 $\mu$Hz which corresponds to a rotation period of 3.4 $\pm$ 0.2 days \textbf{and is confirmed by the observed p modes oscillations}.
\citep{Appourchaux2008}. A second peak is observed at {\bf twice this} frequency and a third one at a frequency four times larger. The second harmonic is ten times weaker than the fundamental but twice {\bf as} strong {\bf as} the fourth harmonic. We see neither a third harmonic, nor a fifth. Such a result would be obtained with a single and long-li{\bf ved} co-rotating active region \citep{Clarke2003} and is also in agreement with the analysis of the light curve {\bf by} \citet{Mosser2009}.

\subsection{Rotational modulation within our sample}

%In addition to the low-frequency lorentzian profile, a super-imposed narrow peak (with at least an harmonics) is a signature of activity {\em and} rotation.
%We stress out that
There may be active stars whose power spectra do not show rotational peaks. A first reason would be an axis of rotation too close to the line of sight.
%Second, surface heterogeneities can be too weakly\textbf{ly contrasted} to induce a detectable rotational modulation.
{\bf A second} reason is a rotation period that is too long to be identified in a too low-frequency resolution Fourier power spectrum. {\bf A third reason could be that stars having spots evolving on time scales shorter than the rotation period do not show a clear rotational modulation.} Indeed, our analysis in terms of LFBI shows cool stars with a high excess power at low frequency, but without any clear rotational modulation. However, 6 stars (including HD\,49933) out of 30 from the seismological sub-sample and 22 stars from the exo-planets sub-sample show a rotational modulation. The fractions of identified rotational modulations in the two sub-samples may be biased by their magnitude distributions. The asteroseismology targets are significantly brighter than the exoplanet ones, so that we detect weakly active stars within the former, but only strongly active ones within the latter. We identified two G stars with a period too long (respectively more than 23 days and more than 29 days) to be confirmed with a 60-day time series. We decided not to include these two stars in our sub-sample of stars with a rotational modulation.

The rotational modulation appears to be significantly more frequent among G and, to a lesser extent, F spectral types (see Table~\ref{tab:rotmod}). The low proportion of K and M stars with an observed rotational modulation seems to be at variance with the expected activity of K and M stars \citep[see for instance][]{Berdyugina2005}. However, our sample of K and M stars is too limited and their rotation periods are likely to be too long for a rotational modulation to be observed in a 60-day photometric time-series. Indeed, \citet{Kiraga2007} identify several M stars with 30-day and longer rotation periods.

\begin{table}
\centerline{\begin{tabular}{|c|c|c|c|}
 \hline spectral  & number  & out of which with  & proportion with  \\
type  &  of stars & \textbf{detected} rotational  & \textbf{detected} rotational \\
& step 1 & modulation & modulation \\
\hline  O/B  & 9 & 0 & 0\% \\
  A & 129 & 2 & 2 \%\\
 F & 140 & 13 & 9 \% \\
 G & 57 & 12 & 21 \%\\
 K/M & 24 & 1  &   4 \%\\
\hline
 Total & 359  & 28  & 8 \%\\
\hline
\end{tabular}}
\caption{Number of selected light curves with a detected rotational modulation signature. The spectral type is derived from a color temperature determined from ground-based photometric observations.}
\label{tab:rotmod}
\end{table}

\begin{figure}
\centering
\includegraphics[height=6cm]{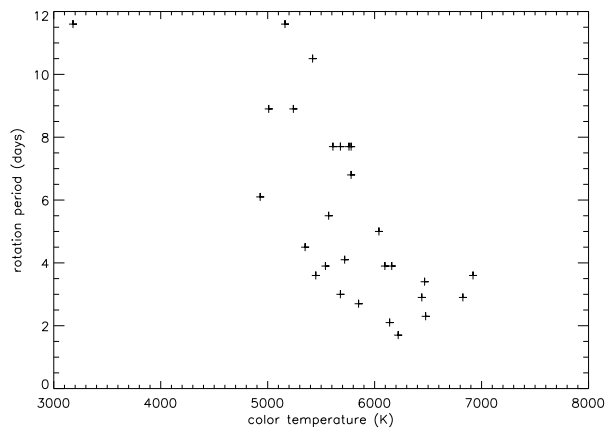}
\caption{Rotation periods versus the color temperature of the 28 stars whose light curves show a rotational modulation.}

\label{rotation}
\end{figure}

The rotation periods of stars with an activity signature are derived from the rotational peaks. In most of the observed cases, the frequency resolution is too low to determine the width of the peak that is attributed to the rotational modulation. In two cases, we can only determine a lower estimate of the rotation period, so that the corresponding stars are excluded from the next steps of our analysis. When plotted against the color temperature (see Fig. \ref{rotation}), the rotation period shows a clear\textbf{ly} decreasing trend \textbf{versus} the color temperature. This result is consistent with previous observations \citep{Gray1982,Lockwood1984}. A significant scatter results from additional parameters which are thought to impact the rotation period (the age for example).

\section{Rossby number}
\label{sec:Rossby}

\subsection{A determinant of stellar activity?}

\textbf{To exhibit solar-like activity, a star needs to have a convective envelope. According to our present understanding of stellar interiors, stars redder (cooler) than F2 have a convective envelope whose thickness decreases withan increasing surface temperature. A dynamo effect is expected to take place when the dynamo number, which compares advective to diffusive effects on the magnetic fields, is larger than 1. Under some very simplified assumptions, a dynamo number can be expressed as a decreasing function of the Rossby number \citep{Parker1979}.}

\textbf{The Rossby number is the ratio between the surface rotation period and the convective turnover time at the bottom of the convective envelope. \citet{Gilliland1985} actually observed that a chromospheric activity index, based on the Ca\,II\,H and K emission fluxes, is correlated with an estimated Rossby number within a 41-star sample. The convective turnover time is expected to be a decreasing function of the effective temperature. A rough relationship between the B-V index and the rotation period can be inferred for main-sequence stars from a statistical study of a large population of stars \citep{Gray1982}. On average, a cooler star rotates more slowly than a hotter one. However, the rotation period and the convective turnover time are likely to show significant scatter within a given spectral type, so that the B-V index is not enough to determine if a star is active or not. The Rossby number is expected to be relevant as an activity index.}

\textbf{From a 277-star sample for which ground-based photometric observations are available, \citet{Hall1991} concluded that the Rossby number of active stars must be lower than 2/3. \citet{Stepien1994} showed that an activity index is well correlated with the Rossby number for a limited color class of main-sequence stars, but he also stressed that, for the other stars, this dimensionless number does not explain activity better than the rotation period. As noted by \citet{Hall1991}, the conditions for a star to be active have generally been addressed focusing on samples in which (very) active stars are over-weighted. A more systematic study of a larger sample of stars, as made possible with CoRoT, spread over the HR diagram could avoid such bias.}

\subsection{Estimating a Rossby number}

The Rossby number is defined as the ratio of the rotation period to the convective turnover time at the bottom of the convective envelope ($\tau_{CV}$). The rotation period is derived from the frequency of the main peak in the Fourier power spectrum.
%We have no observational proxy for the convective turnover time at the bottom of the convective envelope.
Looking at M stars, \citet{Kiraga2007} suggested an empirical determination of the convective turnover time based on a linear relation between the logarithm of an X ray-flux index\textbf{, not available for our targets,} and the rotation period.
%We considered that our sample is too small to fit a color-dependent relation between the rotation period and the activity index as \citet{Stepien1994} did.
\citet{Noyes1984} used a scaling relation between a color index and a convective turnover time.

Here, we prefer to estimate this turnover time using a grid of stellar models with 1 to 2 solar masses, for a star whose effective temperature is {\bf presumed} to be known $\pm$ 100\,K. The global 1D models are obtained with the CESAM code \citep[version 4, see][]{Morel2008} assuming standard physics. Convection is described according to \citet{Bohm58}'s local mixing-length  theory of convection (MLT) with a mixing-length $\Lambda= \alpha_c \, H_p$, where $H_p$ is the pressure scale height and $\alpha_c$ is the mixing-length parameter. The calibration of the associated solar model gives $\alpha_c = 1.62$. Turbulent pressure and microscopic diffusion are not included.
%Overshooting of the convective core is described as an extension of $r_{zc}$ over a distance taken to be min($r_zc,~\alpha_{\rm ov}~H_p$) where $\alpha_{\rm ov}=0.2$ is the overshooting parameter.
All models have a solar iron-to-hydrogen abundance and the chemical mixture of the heavy elements of \citet{GN93}. We use OPAL opacities \citep{Iglesias96} {\bf extended with the} \citet{Alexander94} data for $T\lesssim10^4$\,K, both sets of data being given for {\bf the} \citet{GN93} solar mixture. Finally, the CEFF \citep{JCD92} equation of state is assumed. The convective time $\tau_{CV}$ is then directly deduced from the mixing length
$\Lambda$ and the convective velocity $u$ at one height scale from the base of the convective zone. There is indeed a difficulty in estimating a turnover time at the bottom of the convective envelope where the velocity is zero, {\bf which is why we estimate it} at a height $H_p$ above the bottom.\\
The Rossby number and the low-frequency background index of the Sun are useful references. The solar convective turnover time is computed with the same grid of stellar interior model. We obtain a Rossby number close to 1. The \textbf{LFBI for the Sun} %for a sample of 22 stars
is derived from VIRGO data {\bf for} a 180-day active period. We obtain  \textbf{$log(LFBI)\sim 5$,} which places it among very moderately active stars. 

\subsection{Rossby number and microvariability index}

\begin{figure}
\centering
\includegraphics[height=6cm]{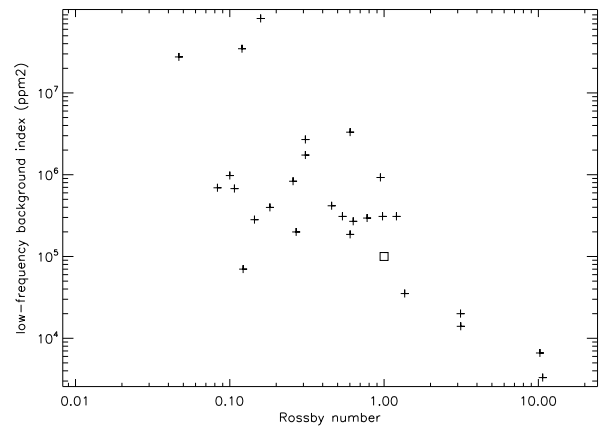}
\caption{The low-frequency background, or microvariability, index versus the Rossby number. The active Sun, i.e. at its maximum, corresponds to the square, while the quiet Sun would have \textbf{an index one hundred times lower (see section 3.3)}.}
\label{rossby}
\end{figure}

In our sample, one star was expected to be fully convective so that our estimation concerns 27 stars from our 28 stars with a rotational modulation. \textbf{Fig.~\ref{rossby}} shows a clear anti-correlation between the activity index and the Rossby number:
%Most of the stars with a detected rotational modulation have a Rossby number which is lower than 1.
%The activity index is decreasing with the Rossby number (see Fig.~\ref{rossby}). We observe less active stars on the asteroseismology channel, i.e. among the brightest stars from our sample.
the most active stars \textbf{having} the lower estimated Rossby number.
\textbf{The observed scatter could result from various parameters, one of which is the stars metallicity. A similar relation was searched for between the rotation period and the activity index. No trend was observed. The Rossby number seems then to be a more useful predictor of the low-frequency microvariability than the rotation period.}
The Rossby numbers are \textbf{limited} by the \textbf{observed} rotational period\textbf{s} and by the color temperature of the selected stars. Based on the CoRoT photometric data from the initial run, our analysis method allows the determination of the rotation period over an interval from a few to nearly 30 days. The upper limit is quite restrictive for the coolest stars. It \textbf{will} be enhanced by using the CoRoT data from longer observation periods. On the other hand, the spectral range studied corresponds to a wide range of convective turnover times (from 0.05 day to 170 days). 
%The convective turnover time is strongly sensitive to the color temperature when considering stars with a shallow convective envelope. Steep variations of the convective turnover time are expected for the coolest stars \citep{Kiraga2007}. Additional observations for this kind of stars could be useful to improve stellar structure models.

\section{Characteristic evolution time of the surface heterogeneities}
\label{sec:evtime}

\subsection{Low frequency power density and characteristic evolution time}

While stellar activity has been mainly addressed in terms of activity index (in Ca\,II\,H and K flux, X{\bf-ray} flux or broadband photometric flux) and activity cycle, based on long observations, there {\bf is} much less available data about the characteristic evolution time of the surface heterogeneities. The parameter $\tau=1/\Delta f$ determined when fitting the Fourier power spectrum to a generalized semi-Lorentzian has been \textbf{previously} used to compute an activity index. The parameter itself gives interesting information about the characteristic evolution time of the surface heterogeneities. 

When there are two different phenomena, for instance spots and faculae, or when spots evolve on a time scale \textbf{that} is shorter than the evolution time of active areas, the lorentzian contribution may be significantly wider than the peaks resulting from the rotational modulation. Such a result is observed in most of the identified stars with rotational modulation. For instance, in the case of HD 49933, $\Delta f$\,=\,2$\mu$Hz corresponds to a 5.5-day characteristic time. This value is slightly {\bf less than} two rotation periods, {\bf whereas} the {\bf width of the} rotational modulation peaks {\bf is} comparable to the frequency resolution {\bf of} a 60-day time series. The rotational modulation could be caused by a phenomenon whose evolution time is quite long compared to the rotation period, \textbf{for example faculae}. 
%The difference between $\Delta f$ and the width of the rotation peak could also result from active longitudes \citep{Berdyugina2005} where long-standing active regions made of spots with a quite short evolution time are more likely to appear. \textbf{However, the analyzed photometric time series are too short to confirm such active longitudes.}

\subsection{The observed characteristic evolution times}

\begin{figure}
\centering
\includegraphics[height=6cm]{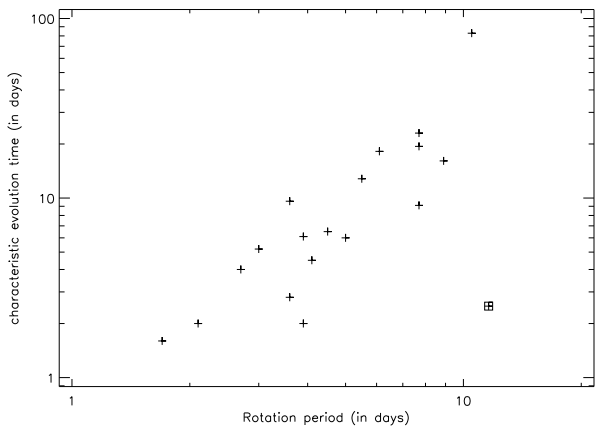}
\caption{The characteristic time $\tau$ against the rotation period. showing a clear increasing trend of $\tau$ with $P_{\rm rot}$. {\bf The slow rotator with a rapid evolution time, indicated by a square, is a M1 star.}
%The quite slow-rotator with a rapid evolution of surface heterogeneities, disclosed with a square, is a M1 star.
}
\label{tau}
\end{figure}

The characteristic time ($\tau$) is plotted against the rotation period ($P$) in Fig.~\ref{tau}, showing a clear increase of  $\tau$ with $P$. Several stars show a rotational modulation caused by a phenomenon whose characteristic time is very large compared to the rotation period.
%Three stars with a high $\tau$ to $P$ ratio have a F5, respectively F9 and G3 spectral type.
In our sample, very few stars show a rotational modulation caused by phenomena with rapid evolution, on a time-scale shorter than the rotation period. {\bf However, there is an interesting exception}, a M1 star with a 11.6-day period and a 2.5-day characteristic evolution time. M stars have a deep convective envelope or are fully convective. Even if they are generally slow rotators, they are expected to be {\bf very} active, as confirmed by X-{\bf ray} flux observations \citep{Kiraga2007}. \textbf{A more extensive sample might allow us to show several different relationships between rotation period and characteristic evolution time. It might be explained by different physical mechanisms linked to surface heterogeneities.}
%More generally, a larger sample of stars is needed to describe more precisely the relation between the characteristic evolution time, the rotation period and the color temperature (or any parameter that allows to position a main-sequence star in the HR diagram).

\section{Conclusion}
\label{sec:conclu}

\subsection{Summary and discussion of our results}

Our first goal was to test an analysis method designed to identify and quantify an activity signature in the photometric data collected by CoRoT. Based on a homogeneous sample of more than 350 stars, we obtain \textbf{ original} results \textbf{regarding a contribution to stellar microvariability from stellar activity}. \textbf{CoRoT, thanks to its sensitivity and stability, allows the investigation of weaker activity stars than from ground-based observations.}

Our work shows that \textbf{CoRoT data allow the detection and measurement in many cases of a low-frequency power excess,} which is interpreted as a photometric signature of stellar activity. After removing several contributions to {\bf the} low-frequency power excess, we are able to fit the low-frequency background to a generalized semi-lorentzian model. A low-frequency background index (LFBI) is then determined in a systematic way. Such an index is more reliable than a microvariability index directly derived from the light curve, because it captures only the low-frequency contribution which is likely to result from stellar activity and excludes other possible sources of variability, {\bf such} as low frequency pulsations.

Despite some scatter, our activity index is correlated to the color temperature, the highest values being obtained for the coolest stars. This result is consistent with previous observations, as reviewed by \citet{Berdyugina2005} {\bf for example}. Our sample allows us to quantify more precisely this relation between the color temperature (or any parameter {\bf locating} a star on the main-sequence) and its level of activity. However, a single-parameter analysis is likely to show significant scatter. Several factors, \textbf{such} as rotation or age and possibly metallicity through its influence on convection, may induce a scatter of an activity index against the temperature. \textbf{Another possible cause for the observed scatter could be the cyclicity of stellar activity: the duration of our observations is likely to be shorter than the activity cycle.}

Our method allows us to quantify an activity index even if there is no rotational modulation. \textbf{However} a \textbf{detected} rotational modulation permits us to go further by estimating the Rossby number. We have seen a clear rotational modulation in the light curves of 28 out of the 430 stars initially included in our sample. There is a higher proportion of such stars within the G, and to a less extent, F spectral types. It could be explained by two biases: hotter stars are less likely to be active and the cooler ones are expected to be slow rotators, so that their rotational modulation is not detected in a 60-day light curve.
%The proportion of stars with such a rotational modulation is quite high. It is more frequent in the G spectral class and, to a more limited extent, in the F type.
The most active stars have a low Rossby number, sometimes {\bf greater} than unity but still anti-correlated with the activity index of the star. This result is consistent with previous work \citep[for instance][]{Hall1991} but our sample allows the exploration of larger values of the Rossby number. Some scatter is also observed, again as expected from a single-parameter analysis. The Rossby number is only a proxy {\bf for} a dynamo number. The scaling relation between the two dimensionless numbers is based on \textbf{the} strong assumption regarding the differential rotation in the convective envelope that may be too approximate for some spectral classes. Moreover, the scatter of the activity index at a given Rossby number can be explained by adverse observing conditions (low activity during a cycle or axis of rotation close to the line of sight). \textbf{In addition, no clear relation is found between activity and rotation period in our sample, confirming the relevance of the Rossby number. However, we also investigated the variation of the characteristic evolution time, which is clearly correlated with the rotation period.} 

Identifying active stars and sorting them according to their physical conditions is an interesting goal in itself.
%The quantification of a characteristic evolution time is a first step on this way. Our sample shows that the characteristic time varies with the rotation period.
%It could be an indication for different activity mechanisms depending on the intensity of the differential rotation. While such a differential rotation seems to be a necessary condition for activity to arise, it could contribute, if it is too intense, to shorten the lifetime of the active regions.

\subsection{Perspectives}

Our sample covers a large part of the HR diagram, but is not unbiased. The exoplanet targets have been mainly selected {\bf among the cooler} main-sequence stars. \textbf{The} available data however cover A to M spectral types. In any case, the study of a larger sample must take into account the non-uniform distribution in terms of color temperature.

In order to study the correlation between an activity index and an estimated Rossby number, we need to determine the rotation period. Some power spectra do not show any low-frequency peaks, but have a high-amplitude low-frequency power excess. The corresponding stars are likely to be active. As was explained for M stars, a power excess at very low frequency can result from an unresolved modulation peak which corresponds to a slow rotation. When an activity index is plotted against the color temperature, the non-rotationally modulated but active stars are taken into account. Unresolved modulation peaks may over-estimate the corresponding activity index.
%A larger sample is here needed to determine more precisely the distribution in the (color temperature, activity index) space.
To study the K and M stars more precisely, longer time-series are needed. More generally, a larger sample of stars is needed to describe more precisely the relation{\bf ship} between the characteristic evolution time, the rotation period and the color temperature (or any parameter that {\bf locates} a main-sequence star in the HR diagram). M stars are again particularly interesting as they are supposed to be fully convective and thus cannot be described \textbf{with the same definition of the} Rossby number. In slightly hotter stars, the convective turnover time is strongly sensitive to the color temperature, since we consider stars with a shallow convective envelope, which are expected to show steep variations of the convective turnover \citep{Kiraga2007}.

CoRoT has now collected numerous longer time series and Kepler is \textbf{collecting} even longer time-series, both missions are thus very promising for the understanding of stellar magnetic activity.

\begin{acknowledgements}

J.-C.H and F.B. thank N. Meunier for helpful comments. 
R.S. thanks E. Michel for providing the stellar models.

\end{acknowledgements}

\bibliographystyle{aa}
\bibliography{projetJCH}

\end{document}